%% file: main.tex
\documentclass[conference]{IEEEtran}
\usepackage{cite}
\usepackage{amsmath,amssymb,amsfonts}
\usepackage{graphicx}
\usepackage{textcomp}
\usepackage{xcolor,soul}
\usepackage{multirow}
\usepackage{comment}
\usepackage[linesnumbered,ruled,vlined]{algorithm2e}
\usepackage[nocenter]{qtree}
\usepackage{tree-dvips}
\usepackage[export]{adjustbox}
\usepackage{listings}
\usepackage{enumitem}
\usepackage{caption,subcaption}
\usepackage{url}

\usepackage{chronology}
\usepackage[linguistics]{forest}
\usepackage{xpatch}
\usepackage{xcolor}
\usepackage{listings}
\usepackage{realboxes}
\usepackage{tikz}

\usepackage [english]{babel}
\usepackage [autostyle, english = american]{csquotes}
\MakeOuterQuote{"}

\begin{document}

\title{Unraveling Log4Shell: Analyzing the Impact and Response to the Log4j Vulnerability}
 
\author{
    \IEEEauthorblockN{
        John Doll,
        Carson McCarthy, 
        Hannah McDougall,
        Suman Bhunia
        }
    \IEEEauthorblockA{Department of Computer Science and Software Engineering, Miami University, Oxford, Ohio, USA}
    
    \IEEEauthorblockA{
    Email: \{dolljm,  mccar130, mcdoughn, bhunias\}@miamioh.edu}

}

\maketitle

\input{abstract}
\input{intro}
\input{background}

\input{methodology}

\input{impact}
\input{defense}

\input{conclusion}

\bibliographystyle{ieeetr}
\bibliography{references}
\end{document}

%% file: abstract.tex
\begin{abstract}

The realm of technology frequently confronts threats posed by adversaries exploiting loopholes in programs. Among these, the Log4Shell vulnerability in the Log4j library stands out due to its widespread impact. Log4j, a prevalent software library for log recording, is integrated into millions of devices worldwide. The Log4Shell vulnerability facilitates remote code execution with relative ease. Its combination with the extensive utilization of Log4j marks it as one of the most dangerous vulnerabilities discovered to date. The severity of this vulnerability, which quickly escalated into a media frenzy, prompted swift action within the industry, thereby mitigating potential extensive damage. This rapid response was crucial, as the consequences could have been significantly more severe if the vulnerability had been exploited by adversaries prior to its public disclosure.

This paper details the discovery of the Log4Shell vulnerability and its potential for exploitation. It examines the vulnerability's impact on various stakeholders, including governments, the Apache Software Foundation (which manages the Log4j library), and companies affected by it. The paper also describes strategies for defending against Log4Shell in several scenarios. While numerous Log4j users acted promptly to safeguard their systems, the vulnerability remains a persistent threat until all vulnerable instances of the library are adequately protected.

\end{abstract}


%% file: intro.tex
\section{Introduction}\label{sec:intro}
The Log4j vulnerability, emerging publicly in December 2021, represents a significant security challenge in recent technological history. Upon its disclosure, security experts rapidly mobilized to mitigate its risks. This vulnerability was given the utmost priority due to its extensive scope and potential impact, earning a maximum severity rating of 10.0 on the Common Vulnerability Scoring System (CVSS) \cite{csrb}. Predominantly integrated into Java applications, Log4j is a staple in hundreds of millions of devices, underscoring the severity of this security lapse \cite{wsj}. In the initial days following its public revelation, there were reports of the vulnerability being exploited approximately 2 million times per hour \cite{zdnet}. The breach allowed attackers to execute arbitrary code on targeted systems by manipulating log messages or their parameters. Highlighting the ease of exploitation, a demonstration on \emph{YouTube} showed the vulnerability being used to gain control over another player's computer in the widely popular game \emph{Minecraft} \cite{minecraft} \cite{youtube}.

The attack commenced when a malicious actor initiated a Java Naming and Directory Interface (JNDI) query to the target server. The JNDI library, central to this vulnerability, allowed the exploit to function, though it was also operable with other types of requests \cite{alertlogic}. The critical element of this attack involved transmitting a string embedded with a link to the malicious payload via the query, leading to the execution of arbitrary code on the target system \cite{mitre}. This capability effectively granted the attacker remote control over the server. Subsequently, the attacker could establish a foothold on the server, setting the stage for more advanced and potentially damaging attacks.

Security professionals were compelled to respond swiftly to the vulnerability to mitigate the risk of widespread attacks, given the pervasiveness and simplicity of the vulnerability. The vulnerability's nature also enabled automated reconnaissance by malicious actors, further escalating the attack frequency \cite{csrb}. Despite the rapid deployment of firewalls and other protective measures, adversaries found ways to circumvent these defenses and continue exploiting the vulnerability. A patch was released two weeks after the development team became aware of the vulnerability \cite{loggingServices}, and organizations promptly began its implementation \cite{cisa}. Fortunately, despite these challenging conditions, few major breaches were reported \cite{standaard}.

The Log4j vulnerability has prompted several defense recommendations. These include auditing applications for Log4j library usage in handling user inputs, identifying additional instances of Log4j utilization, and upgrading to a Log4j version compatible with the organization's Java version \cite{cisa}\cite{upguard}. If updating to a secure Log4j version is infeasible, alternative mitigation measures can be employed, or the Jndilookup.class file can be removed from the Log4j library \cite{loggingServices}. However, while these measures can prevent future attacks, they cannot reverse the damage already inflicted.

The simplicity of the Log4j vulnerability was notably alarming. Prior to defensive efforts, even inexperienced hackers, following instructions from a \emph{YouTube} video, could execute the hack. The exploit saw immediate and widespread use, ranging from government-backed intelligence groups to novice hackers. The ubiquitous presence of the Log4j library in hundreds of millions of devices worldwide, coupled with its open-source, efficient nature, raises critical questions about the oversight in security practices that allowed such a significant vulnerability to remain undetected in widely-used software for an extended period.

In this paper, we will explore the background of the Log4j vulnerability, detailing the nature of Log4j, the timeline of its discovery, and the subsequent weaponization of the vulnerability. We will provide an in-depth analysis of the attack process, elucidating how Log4j functions as a gateway for more sophisticated attacks on target computers. The vulnerability inherent in Log4j facilitated remote server control by actors, enabling them to execute arbitrary scripts. Additionally, the paper will assess the impact of this vulnerability, focusing on how prompt actions by security professionals played a crucial role in averting major incidents stemming from the exploit.

In summary, the main contributions of this paper are:
\begin{itemize}

    \item \textbf{Detailed Analysis of Log4Shell Vulnerability}: Offers an in-depth look at the Log4Shell vulnerability in the Log4j library, encompassing its discovery, exploitation methods, and impact across various sectors.
    
    \item \textbf{Mitigation and Defense Strategies}: Presents a range of strategies for mitigating the vulnerability, tailored for different scenarios, including environments where updating Log4j is not feasible.
    
    \item \textbf{Insights into Open Source Software Security}: Raises important discussions about the security challenges in open-source software, using the Log4j incident to highlight the need for enhanced security practices in open-source development and maintenance.
    
\end{itemize}

%% file: background.tex
\section{Background}\label{sec:background}

\subsection{Introduction}

This section provides an in-depth examination of Log4j and Log4Shell, highlighting their interrelation and distinct aspects. It delves into the discovery process of Log4Shell and the subsequent steps undertaken post-discovery. Finally, it details the immediate reactions and measures implemented by community members and governmental bodies in response to the vulnerability.

\subsection{Log4j and Log4Shell Introduction}

Log4j, a Java-based software library, is integral to logging and monitoring activities in applications and servers. Developed under the Apache Logging Services, a project of the Apache Software Foundation \cite{loggingServices}, its vulnerability became widely known in December 2021. Dubbed Log4Shell, this vulnerability first manifested in \emph{Minecraft: Java Edition} \cite{minecraft} and enabled malicious actors to execute code on targeted clients or servers. Typically resulting in a reverse shell, this exploit facilitated unauthorized access to servers, hence the vulnerability's moniker, Log4Shell. Often the initial phase in a broader attack strategy, Log4Shell was pivotal in breaching target systems. Its risk factor was assessed at the maximum score of 10 on the Common Vulnerability Scoring System (CVSS), reflecting its exploitation ease and Log4j's ubiquity in various applications \cite{mitre}.

\subsection{Discovery of Log4Shell}

\par On November 24, 2021, an Alibaba security engineer uncovered the Log4Shell vulnerability using recursive analysis. This discovery revealed the possibility of remote code execution on different machines \cite{arstechnica}. The vulnerability, along with a proof of concept, was promptly reported to the Apache Software Foundation. This marked the prelude to the vulnerability's public acknowledgment. Although the ASF initiated work on a resolution by November 29, a security advisory wasn't issued immediately \cite{csrb}. Subsequent modifications to the Log4j library on GitHub were made on December 5, followed by a new Log4j release the next day \cite{csrb}. On December 9, the same engineer alerted ASF about discussions on exploiting the vulnerability on WeChat \cite{csrb}.

\subsection{Immediate Impact from Log4Shell}

\par A \emph{YouTube} video demonstrating the exploit in Minecraft underscored the severity of the vulnerability, affecting Minecraft versions 1.8.8 to 1.18 \cite{arstechnica}. However, the reach of Log4Shell extended far beyond Minecraft, potentially jeopardizing "hundreds of millions of devices" \cite{wsj}. Cybersecurity firm Akamai reported "around 2M attack attempts per hour" from December 9th to 16th \cite{zdnet}. Following a surge in social media posts about Log4Shell, ASF released a public fix on December 10 and assigned CVE-2021-44228 to the vulnerability. This identifier was later rated a maximum of 10.0 on the CVSS by the NIST National Vulnerability Database (NVD) \cite{csrb}. The CSRB Report provides a comprehensive account of the unfolding events and critical milestones \cite{csrb}.
\input{timeline}

\begin{table}[]
    \centering
    \caption{Overview of Key Log4j Vulnerabilities: Ratings and Affected Versions}
    \label{tab:my_label}
        \scriptsize

\begin{tabular}{|p{0.2\linewidth} | p{0.2\linewidth} | p{0.45\linewidth}|}
    \hline
    Vulnerability & Rating & Versions Effected \\
    \hline
    CVE-2021-44832 & 6.6 &  All versions from 2.0-beta7 to 2.17.0, excluding 2.3.2 and 2.12.4 \\
     \hline
     CVE-2021-45046 & 9.0 & All versions from 2.0-beta9 to 2.15.0, excluding 2.12.2 \\
     \hline 
     CVE-2021-44228 & 10.0 & All versions from 2.0-beta9 through 2.15.0 \\
     \hline
\end{tabular}

\end{table}

%% file: timeline.tex
\newcommand\ytl[2]{
\parbox[b]{1in}{\hfill{\color{cyan}\bfseries\sffamily\small #1}~$\cdots$~}\makebox[0pt][c]{$\bullet$}\vrule\quad \parbox[c]{2in}{\vspace{7pt}\color{red!40!black!80}\raggedright\sffamily\small #2.\\[7pt]}\\[-3pt]}

\begin{figure}
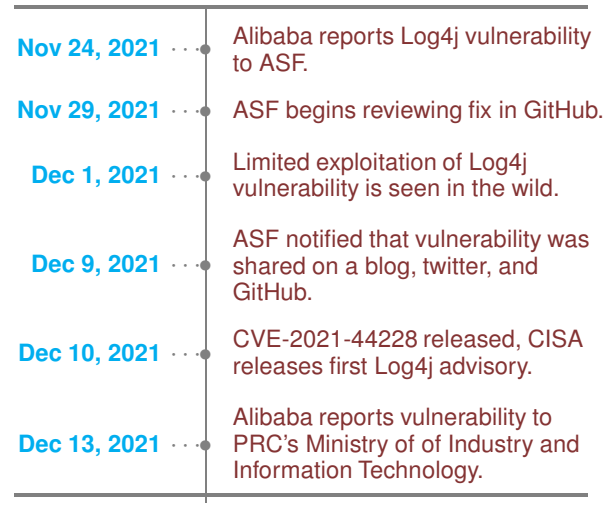

    \centering
    \begin{minipage}{3in}
        \color{gray}
        \rule{\linewidth}{1pt}
        \ytl{Nov 24, 2021}{Alibaba reports Log4j vulnerability to ASF}
        \ytl{Nov 29, 2021}{ASF begins reviewing fix in GitHub}
        \ytl{Dec 1, 2021}{Limited exploitation of Log4j vulnerability is seen in the wild}
        \ytl{Dec 9, 2021}{ASF notified that vulnerability was shared on a blog, twitter, and GitHub}
        \ytl{Dec 10, 2021}{CVE-2021-44228 released, CISA releases first Log4j advisory}
        \ytl{Dec 13, 2021}{Alibaba reports vulnerability to PRC's Ministry of of Industry and Information Technology}
        \bigskip
        \rule{\linewidth}{1pt}%
    \end{minipage}
    
    \caption{Timeline of Log4j Vulnerability \cite{csrb}}
    \label{fig:timeline}
 \end{figure}

%% file: methodology.tex
\section{Attack Methodology}\label{sec:method}

The attack methodology section of this paper aims to elucidate how an attacker could exploit the Log4Shell vulnerability. It is structured to cover the attack in three main phases and discusses the enabling technology behind the exploitation. Subsequently, the section delves into evasion techniques and advancements made to the attacks. The final part of the section addresses how prior knowledge facilitated quicker execution of these attacks.

\subsection{Technology Allowing Exploitation and Attack Phase 1}
The Log4Shell vulnerability originates from insufficient safeguards against "attacker controlled LDAP and other JNDI related endpoints" \cite{mitre}. LDAP, or Lightweight Directory Access Protocol, is an internet protocol for accessing and managing distributed directory services, while JNDI, the Java Naming and Directory Interface, facilitates resource lookup by name. The vulnerability's crux lies in Log4J's message lookup substitution feature, enabling attackers to execute arbitrary code from LDAP servers by controlling log messages or their parameters \cite{mitre}\cite{cisa}. Log4J processes special markers "[]" in log entries, triggering JNDI queries upon request processing. The attack commences with a crafted string like [\textit{jndi: ldap: //exampleattacker.com/a}]. The server, interpreting this marker, sends an LDAP request to the specified attacker-controlled site for a Java class. Upon receiving the response, the server decodes it, unintentionally executing the payload. Hence, a strategically composed string embedding a malicious payload link can commandeer the operating system through the JNDI query. Researchers discovered that the exploit extends beyond LDAP to include DNS, NIS, NDSM RMI, IIOP, and COBRA, setting off the attack's second phase \cite{alertlogic}. The simplicity of this exploit, as noted by the SANS Institute, highlights its alarming potential \cite{rce}.

\subsection{Attack Phase 2}
The second phase of the attack is characterized by the attacker's efforts to establish a foothold on the victim's machine via remote code execution, exploiting the JNDI request initiated in the first phase. This stage involves a trial-and-error process with two primary potential failure points. The first challenge for the attacker is to have access to a server that can receive and process the JNDI-based LDAP request originating from the victim's server. Hackers often rely on compromised servers for this purpose, which may have disabled features, unreliable infrastructure, or could be offline, all of which could result in LDAP connection failure \cite{alertlogic}. The second challenge arises if the LDAP connection is successful; the victim's server must then download and execute the script sent by the attacker. This step can fail if the server lacks necessary components or elements needed for the script's execution. However, failure at this stage does not mark the end of the attack but is merely a temporary obstacle. The attacker can modify the LDAP request in the JNDI query or adjust the script for a subsequent attempt \cite{broadcom}. Successful execution of the script typically grants the attacker a reverse shell, providing a solid base for launching more advanced attacks in phase 3.

\subsection{Attack Phase 3}
The third phase of the attack is divided into two subphases: Staging and Execution. In the Staging subphase, which begins once the malicious script is downloaded and executed on the target machine, the attacker solidifies their foothold. This stage is critical for preparing a more comprehensive attack, involving tasks such as installing additional tools or attempting to elevate credentials \cite{alertlogic}. This preparation serves as the foundation for subsequent actions. During the Execution subphase, the attacker utilizes their newfound system access and tools for malicious purposes. Activities during this phase can range from installing malware or cryptocurrency miners, encrypting the system for a ransomware attack, to exfiltrating data for sale or public release \cite{alertlogic}. This phase mirrors the progression seen in many other cyber attacks following unauthorized system access and is not the central focus of this paper.

\subsection{Improvements on Attack Sophistication}
Attackers rapidly identified and utilized unique methods to exploit the vulnerability. Insights from a report by the U.S. Department of Homeland Security's Cyber Security Review Board included notable trends in the industry. One such trend was the automation of the reconnaissance process using botnets, a technique widely adopted by attack groups to swiftly pinpoint vulnerable targets. In the period from January to March 2022, cybersecurity technology companies recorded and investigated several ransomware attacks that exploited the Log4J vulnerability, underscoring the widespread impact and adaptability of the exploit \cite{csrb}.
\par
\input{attackPhase}
\subsection{Attempts at Evasion Utilizing Log4J}
Another way the attack was iterated and improved upon was through the use of evasion techniques to avoid detection and blocking by firewall or other traffic monitoring rules. Early into the discovery and abuse of the Log4Shell vulnerability, security professional began to create firewall rules to attempt to block traffic that contained strings that appeared to be exploiting Log4Shell. Initial attack attempts were rudimentary and the strings had no obfuscation or evasion attempts. They often took the form previously shown of [\textit{jndi: ldap: //exampleattacker.com/a}]. A simple attack attempt can be halted by simple rules looking for phrases in logged content such as \${jndi:dns...}, \${jndi:rmi...} and \${jndi:ldap...}. Eventually, actors began to make attempts at obfuscation, such as replacing \$\{ with \%24\%7B or \textbackslash u0024\textbackslash u007b \cite{cloudflareEvasion}. However these tricks were not very effective as the logs can easily be decoded before checking for malicious strings. Log4J, however, has a large set of tools that hackers could unfortunately take use of to evade detection. Certain functions such as the {lower} function could be used to hide JNDI requests, as the function returns only the letter and concatenates it with the rest of the log, meaning J, N, D and I could be strung together in {lower} functions and still provide functionality with the following input "\$\{lower:\$\{lower:J\}\}\$\{lower:\$\{upper:D\}\}\$\{lower:N\}i", which when handled by Log4J will return "jndi". A full string could look like: [\textit{\$\{lower:\$\{lower:J\}\}\$\{lower:\$\{upper:D\}\}\$\{lower:N\}i: ldap: //exampleattacker.com/a}] \cite{microsoft} \cite{cloudflareEvasion}. Other Log4J methods were used by hackers to avoid evasion, and professionals had to continuously update the firewall rules to attempt to block this traffic.
\par
\subsection{Previous Knowledge Enabled Faster Attacks}
The exploitation of JNDI/LDAP for remote code execution is not a novel concept. This was evidenced at BlackHat 2016, where Hewlett Packard security experts Alvaro Muñoz and Oleksandr Mirosh showcased the potential for JNDI requests to facilitate remote code execution on servers. They explored three attack vectors: RMI, COBRA, and LDAP. Specifically, the LDAP vector encompassed multiple scenarios, including attacks on vulnerable LDAP servers and applications \cite{blackhat}. The pre-existing knowledge of these vectors and methods significantly contributed to the efficacy of the Log4J vulnerability. The Log4J's configuration, which allowed for handling special markers facilitating JNDI requests, became an ideal mechanism for executing JNDI/LDAP remote code attacks.

%% file: attackPhase.tex
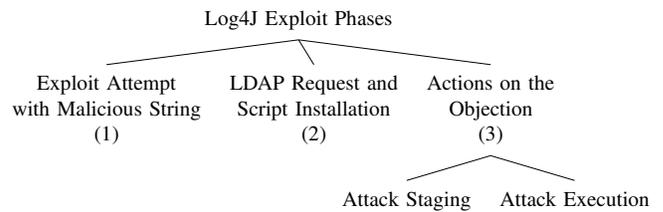
\begin{figure}

 \begin{center}
    \scalebox{0.8}{
     \begin{forest}
         [Log4J Exploit Phases
          [{Exploit Attempt \\ with Malicious String\\(1)}]
          [{LDAP Request and \\ Script Installation\\(2)}]
          [{Actions on the \\ Objection\\(3)}
            [Attack Staging]
            [Attack Execution]
          ]
         ]
     \end{forest}}
 \end{center}
    \caption{Phases of an Attack Using Log4Shell \cite{csrb}}
    \label{fig:attackPhase}
 \end{figure}

%% file: impact.tex
\section{Impact}\label{sec:impact}

\subsection{Introduction}
This section attempts to both quantify and qualify the impact that the Log4Shell vulnerability has had and will continue to have. First, the initial estimates and their impact on the workforce are discussed. Next, the extremely high level of exploitation is shown. Finally, examples of exploitation against both government and enterprise assets are given.

\subsection{Initial Estimates and Response}
As mentioned in the background section, early estimates from the US Government, reported by the Wall Street Journal, indicated that the number of devices vulnerable to the Log4Shell issue was in the hundreds of millions \cite{wsj}. Following the vulnerability's disclosure, sectors such as Information Technology, Security, and Software Engineering entered a state of crisis management. Organizations were tasked with identifying which devices, tools, and software were at risk and devising strategies to monitor these potential points of attack. The Apache Software Foundation responded swiftly, releasing a patch on December 6th, 2021, approximately two weeks after learning of the vulnerability – a notably rapid response \cite{loggingServices}. However, widespread dissemination of information about the vulnerability on platforms like GitHub and social media outpaced the patch's implementation in many systems \cite{csrb}. A significant challenge in addressing this vulnerability was the pervasive use of Log4J; countless tools depended on it, and these tools were, in turn, integrated into other systems. This interdependency created multiple hurdles for organizations attempting to apply patches, often requiring them to wait for other entities to complete their patching processes. To expedite remediation, the CISA mandated that all software assets vulnerable to Log4Shell be updated or removed from agency networks by December 24th, two weeks after the vulnerability was catalogued by the agency \cite{cisa}.
\par 
\subsection{High Level of Exploitation}
After the Log4Shell exploit details were widely disseminated, a surge in its exploitation by hackers and malicious actors was observed. A significant increase in mass scanning for the vulnerability was reported by a security firm within just two hours of the information being posted on GitHub \cite{csrb}. Cloudflare, a well-known content delivery network, detected an average of 400 unique exploit attempts per second within five days of the vulnerability becoming public, amounting to over one million requests every hour. The exact number of exploit attempts is likely indeterminable, as early use of obfuscation techniques, as outlined in the Attack Methodology section, complicated tracking efforts. To counteract this, Cloudflare implemented stringent Web Application Firewall (WAF) rules. Between December 10th and December 13th, an average of 19,042.5 requests were blocked per minute, totaling 27,421,200 blocks over the four-day period \cite{cloudflareEvasion}. While some legitimate traffic might have been inadvertently blocked, the sheer volume of blocked requests underscores the extensive efforts to exploit this vulnerability in its early stages. This unprecedented level of exploitation led to Log4Shell being ranked at the top of CISA's "Top 15 Routinely Exploited Vulnerabilities" for 2021, despite the exploit only emerging in December of that year \cite{cisaTopVuln}.

\par
\subsection{Attacks on Government Resources}
Despite the high volume of exploit attempts following the discovery of the Log4Shell vulnerability, reports of major breaches were relatively few. One of the first significant attacks targeted the Belgian Ministry of Defense. Reported on December 12th, just three days after the vulnerability became widely known, the Ministry did not disclose specific details but confirmed the Log4J vulnerability as the attack vector. They implemented quarantine measures to isolate the affected network \cite{standaard}. 

Microsoft reported that cyberattack groups from China, Iran, North Korea, and Turkey utilized Log4Shell for attacks throughout December 2021 and into January 2022. Notably, APT41, a group from China, began exploiting Log4Shell mere hours after the Apache Software Foundation issued its public warning. Cybersecurity firm Mandiant revealed that APT41 leveraged the vulnerability in a campaign against multiple US state governments, first detected in May 2021 and persisting until February 2022. At least six state governments were compromised. The campaign initially used the USAHerds exploit but later incorporated Log4Shell for reconnaissance and to install backdoors for gaining a foothold in various systems \cite{apt41}. 

These instances are just a few among numerous attacks targeting national and state governments. Additionally, many corporations also fell victim to the Log4Shell vulnerability.

\par
\subsection{Vulnerabilities Within Enterprise Software}
Following the public disclosure of Log4Shell, numerous companies rapidly commenced evaluations to identify tools and software susceptible to the vulnerability. VMware was among the companies significantly impacted, identifying as many as 56 products vulnerable to the exploit. In an advisory issued on December 10th, VMware outlined these vulnerabilities. They subsequently updated the advisory on the next day with workarounds for multiple software products and continued to provide updates as patches were developed and applied \cite{vmwareVuln}.

VMware Horizons, a desktop and application virtualization platform, emerged as a prime target for attacks. Investigations by Sophos, a British IT security company, revealed that many Horizon servers were compromised from December 2021 into January 2022. Attackers exploited Log4Shell to alter legitimate Java files on the servers, installing web shells that facilitated remote code execution. These web shells were frequently used to download cryptocurrency mining malware, further exacerbating the impact of the attacks \cite{vmwareHorizon}.

%% file: defense.tex
\section{Defense Solution}\label{sec:defense}

\lstset{
  basicstyle=\ttfamily,
}

\subsection{Introduction}

The defense section goes through the multiple different ways to help guard from the Log4Shell vulnerability. It offers different scenarios that might require different routes to a solution to be taken. Specific versions of the library are discussed as well as the actions that need to be taken depending on the state of the given software.

\subsection{Upgrading Log4j Package}

Defending against Log4Shell involves several key steps. The Cybersecurity and Infrastructure Security Agency (CISA) recommends first identifying all internet-facing assets that accept user input and utilize the Log4j library \cite{cisa}. These assets are particularly critical and vulnerable as they are accessible to the public and easily exploited. The next step involves locating any other instances where the Log4j library is used, which can be achieved using scanners available on CISA's and Carnegie Mellon's CERT GitHub repositories \cite{cisa}.

Once all instances of Log4j usage are identified, upgrading to the latest Log4j version is the most effective defense. For Java 8, upgrade to at least Log4j version 2.17.1; for Java 7, use version 2.12.4; and for Java 6, version 2.3.2 is recommended \cite{loggingServices} \cite{cisa} \cite{rce} \cite{comptia} \cite{gsa} \cite{ftc} \cite{ncsc}. These updated versions are available on the Apache website. Post-upgrade, dependencies should be updated to reference the new, secure Log4j versions. While upgrading prevents future attacks, it does not rectify any damage already inflicted by previous attacks \cite{upguard}.

The need for these specific version upgrades stems from vulnerabilities identified in earlier patches. Version 2.15.0's fix, intended to restrict JNDI lookups to localhost, was found vulnerable under non-default configurations \cite{loggingServices}. Attackers could exploit Context Lookups in non-default Pattern Layouts to manipulate the Thread Context Map (TMC) \cite{loggingServices}. Subsequent versions, 2.16.0 and 2.17.0, had their own vulnerabilities, including susceptibility to Denial of Service (DoS) attacks and stack overflow issues due to lack of recursion detection \cite{loggingServices}. At the time of writing, versions 2.17.2, 2.18.0, and 2.19.0 are free of known vulnerabilities, with 2.19.0 being the latest version.

\tikzstyle{decision} = [rectangle, minimum width=1cm, minimum height=0.3cm, text centered, draw=black, fill=pink!30]

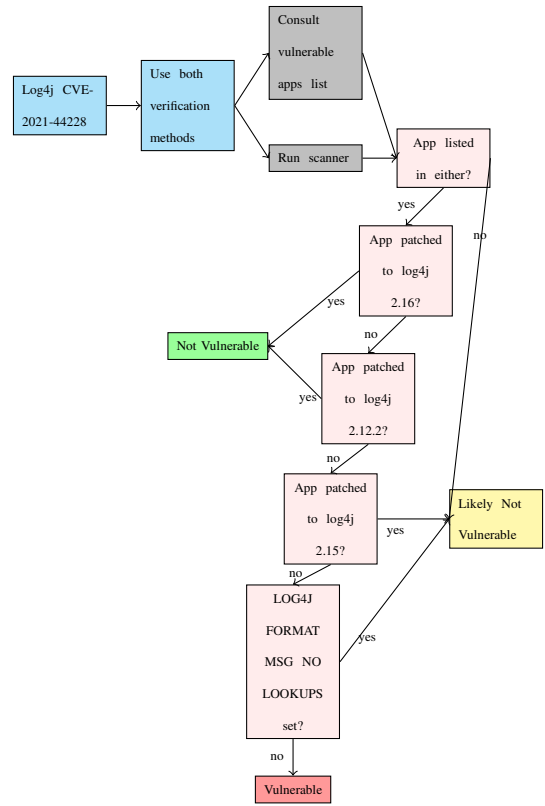
\begin{figure}
    \centering
\begin{tikzpicture} 
    \node (issue) [fill=cyan!30, text width=1cm, draw=black] {\tiny Log4j CVE-2021-44228};
    \node (twoVer) [fill=cyan!30, right of=issue, xshift=0.7cm, text width=1cm, draw=black] {\tiny Use both verification methods};
    
    \node (app) [fill=black!25, right of=twoVer, xshift=0.7cm, yshift=0.7cm, draw=black, text width=1cm] {\tiny Consult vulnerable apps list};
    \node (scanner) [fill=black!25, right of=twoVer, xshift=0.7cm, yshift=-0.7cm, draw=black, text width=1cm] {\tiny Run scanner};
    
    \node (listed) [decision,  right of=scanner, xshift=0.7cm, text width=1cm] {\tiny App listed in either?};
    \node (patched1) [decision, below of=listed, yshift=-0.5cm, xshift=-0.5cm, text width=1cm] {\tiny App patched to log4j 2.16?};
    \node (patched2) [decision, below of=patched1, yshift=-0.7cm, xshift=-0.5cm, text width=1cm] {\tiny App patched to log4j 2.12.2?};
    \node (patched3) [decision, below of=patched2, yshift=-0.6cm, xshift=-0.5cm, text width=1cm] {\tiny App patched to log4j 2.15?};
    \node (set) [decision, below of=patched3, yshift=-0.9cm, xshift=-0.5cm, text width=1cm] {\tiny LOG4J FORMAT MSG NO LOOKUPS set?};
    
    \node(vulnerable) [below of=set, draw=black, fill=red!40, yshift=-0.7cm] {\tiny Vulnerable};
    \node(notVulnerable) [left of=patched2, xshift=-1cm, yshift=0.7cm, draw=black, fill=green!40] {\tiny Not Vulnerable};
    \node(likelyNot) [right of=patched3, xshift=1.2cm, draw=black, fill=yellow!40, text width=1cm] {\tiny Likely Not Vulnerable};
    
    \draw[->] (issue.east) -- (twoVer.west);
    \draw[->] (twoVer.east) -- (scanner.west);
    \draw[->] (twoVer.east) -- (app.west);
    \draw[->] (scanner.east) -- (listed.west);
    \draw[->] (app.east) -- (listed.west);
    \draw[->] (listed.south) -- node[anchor=east] {\tiny yes} (patched1.north);
    \draw[->] (patched1.south) -- node[anchor=east] {\tiny no}(patched2.north);
    \draw[->] (patched2.south) -- node[anchor=east] {\tiny no}(patched3.north);
    \draw[->] (patched3.south) -- node[anchor=east] {\tiny no}(set.north);
    \draw[->] (set.south) -- node[anchor=east] {\tiny no}(vulnerable.north);
    \draw[->] (patched1.west) -- node[anchor=north, , near start] {\tiny yes}(notVulnerable.east);
    \draw[->] (patched2.west) -- node[anchor=north, near start] {\tiny yes}(notVulnerable.east);
    \draw[->] (listed.east) -- node[anchor=south, near start] {\tiny no} (likelyNot.west);
    \draw[->] (patched3.east) -- node[anchor=north, near start] {\tiny yes}(likelyNot.west);
    \draw[->] (set.east) -- node[anchor=north, near start] {\tiny yes} (likelyNot.west);
\end{tikzpicture}
    
    \caption{Decision Flowchart for Assessing Vulnerability to Log4j}
    \label{fig:enter-label}
\end{figure}

\subsection{Mitigation with Vulnerable Log4j Package}

In cases where updating Log4j is not feasible, there are alternative mitigation strategies to address the Log4Shell vulnerability. The Cybersecurity and Infrastructure Security Agency (CISA) outlines five such tactics:

\begin{enumerate}
    \item \textbf{Physical Removal from Network}: Physically disconnect the device from the network. This can be done by unplugging the device to ensure it is no longer powered or connected to the internet.
    \item \textbf{Isolation in a 'Jail VLAN'}: Segregate affected devices into a 'jail VLAN' with enhanced monitoring and security measures.
    \item \textbf{Network Layer Traffic Blocking}: Implement traffic control at the network layer using devices like switches to block data flow to and from the vulnerable system.
    \item \textbf{Firewall with Stringent Port Control and Logging}: Establish a firewall with strict port control and detailed logging to allow only authorized traffic and provide a record of connection attempts.
    \item \textbf{Restricting Communication with Affected Assets}: Limit the communication of an affected network asset with the internet and the enterprise network to prevent the establishment of a shell or, if a shell is present, to isolate the device and protect the network.
\end{enumerate}

These measures, while not a replacement for updating to a secure version of Log4j, can significantly reduce the risk of exploitation from the Log4Shell vulnerability.

\subsection{Editing a Vulnerable Log4j Package}

If updating Log4j to the latest version isn't feasible, but access to modify the Log4j library is available, there are specific measures that can be taken to enhance security, particularly concerning the `\lstinline[language=TeX]|Jndilookup.class|` file. The safest option is to remove the \lstinline[language=TeX]|Jndilookup.class| from the class path, which can be done by executing the following command: \lstinline[language=TeX]|zip -q -d log4j-core- *.jar org /apache/logging/log4j|
\lstinline[language=TeX]|/core/lookup/JndiLookup.class| \cite{upguard} \cite{mitigation}. In an article on \emph{TowardsDev}, the author adds that changing the system property "\lstinline[language=TeX]|log4j2.formatMsgNoLookups|" to "\lstinline[language=TeX]|true|" will also suffice for stopping the Log4Shell attack, although this method has been found to leave the door open to a separate attack using Thread Context \cite{loggingServices} \cite{towardsDev}. If the \lstinline[language=TeX]|Jndilookup.class| is still required by the software making use of the Log4j library, a second approach is to delete or rename the \lstinline[language=TeX]|Jndilookup.class|. CISA notes that in doing so, "removal of the \lstinline[language=TeX]|JndiManager| will cause the \lstinline[language=TeX]|JndiContextSelector| and \lstinline[language=TeX]|JMSAppender| to no longer function" \cite{mitigation}. There are also several hot patches that have been released and implemented that are meant just as short term fixes until the Log4j software can be fully updated to a safe version.

\subsection{Continued Mitigation Tactics}

Alongside upgrading the Log4j library or implementing alternative mitigation strategies, regular monitoring and analysis of logs and systems are crucial to ensure that both new and existing software remain secure. In the initial stages of the Log4Shell vulnerability's emergence, it was even recommended to establish a dedicated team to address this specific issue \cite{microsoft}. Regularly scheduled scans of the network and source code are essential to confirm that no software updates are overlooked \cite{mitigation}. Tools like Microsoft's Defender for Endpoint are designed for such purposes, providing system administrators with reports highlighting potential vulnerabilities \cite{microsoft}.

The widespread impact of Log4Shell has reignited discussions in the industry about the importance of automated software update tools, which can ensure that software packages within systems are promptly updated when new versions are released \cite{comptia}. It is also important to communicate with any third parties connected to a codebase that uses a vulnerable version of Log4j. These parties should be informed of the updates and provided with guidance on securing their systems \cite{ftc} \cite{ncsc} \cite{mitigation}. For a comprehensive list of recommended actions in response to the Log4j vulnerability, the CISA’s Alert (AA21-356A) offers a detailed overview \cite{mitigation}.

%% file: conclusion.tex
\section{Conclusion}\label{sec:conclusion}
The Log4j vulnerability, emerging in December 2021, represented a significant shock to security professionals due to its simplicity and the extent of its potential impact. Immediately after becoming public knowledge, the vulnerability was exploited in millions of attacks worldwide. It provided hackers with the ability to remotely execute code on targeted machines, granting them unrestrained access. The range of exploiters spanned from government-backed offensive hacking groups to amateur hackers, often referred to as "script kiddies."

The critical nature of this vulnerability prompted developers to swiftly release patches and establish mitigation procedures. These efforts proved largely effective in preventing major attacks on corporations and governments. However, the challenge remains immense, given the widespread usage of Log4j across millions of systems globally. Ensuring that all these systems are updated is a daunting yet crucial task.

Additionally, the incident has sparked broader discussions about the reliance on open-source software, which is often maintained by volunteers. The presence of such a significant yet undetected vulnerability in Log4j for years raises concerns about the overall security and oversight of open-source software projects, highlighting the need for more rigorous security practices in their development and maintenance.